\documentclass[nofootinbib,superscriptaddress,
showpacs]{revtex4}
\usepackage{graphicx}
\begin{document}

\def\d{{\rm d}}
\def\la{\langle}
\def\ra{\rangle}
\def\om{\omega}
\def\Om{\Omega}
\def\vep{\varepsilon}
\def\wh{\widehat}
\def\tr{\rm{Tr}}
\def\da{\dagger}
\newcommand{\beq}{\begin{equation}}
\newcommand{\eeq}{\end{equation}}
\newcommand{\beqa}{\begin{eqnarray}}
\newcommand{\eeqa}{\end{eqnarray}}
\newcommand{\intf}{\int_{-\infty}^\infty}
\newcommand{\into}{\int_0^\infty}
\newcommand{\HC}{H_{\mathrm{c}}}             
\renewcommand{\L}{\mathcal{L}}               
\newcommand{\R}{\mathcal{R}}
\renewcommand{\P}{\mathcal{P}}
\newcommand{\Q}{\mathcal{Q}}                 
\newcommand{\ket}[1]{|#1\rangle}             
\newcommand{\bra}[1]{\langle #1|}            
\newcommand{\braket}[2]{\langle #1|#2\rangle}
\renewcommand{\d}[1]{\mathrm{d}#1}           
\newcommand{\scr}[1]{\mathscr{#1}}           
\renewcommand{\v}[1]{\mathbf{#1}}            
\renewcommand{\Re}{\mathrm{Re}}              
\renewcommand{\Im}{\mathrm{Im}}              
\newcommand{\hc}{\mathrm{h.c.}}              
\newcommand{\mt}[1]{\mathrm{#1}}             
\newcommand{\bgl}[1]{\boldsymbol{#1}}         

\title{Seeking better times: atomic clocks in the generalized Tonks-Girardeau
regime}
\author{S. V. Mousavi}
\email{s_v_moosavi@mehr.sharif.edu}
\address{Department of Physics, Sharif University of Technology, P. O. Box 11365-9161, Tehran, Iran}
\author{A. del Campo}
\email{adolfo.delcampo@ehu.es}
\address{Departamento de Qu\'imica-F\'isica,
UPV-EHU, Apartado. 644, Bilbao, Spain}
\author{I. Lizuain}
\email{ion.lizuain@ehu.es}
\address{Departamento de Qu\'imica-F\'isica,
UPV-EHU, Apartado. 644, Bilbao, Spain}
\author{M. Pons}
\email{marisa.pons@ehu.es}
\address{Departamento de F\'isica Aplicada I,
E.U.I.T. de Minas y Obras P\'ublicas, UPV-EHU, 48901 Barakaldo, Spain}
\author{J. G. Muga}
\email{jg.muga@ehu.es}
\address{Departamento de Qu\'imica-F\'isica,
UPV-EHU, Apartado. 644, Bilbao, Spain}

\begin{abstract}
First we discuss briefly the importance of time and time keeping, 
explaining the basic functioning of clocks in general and of atomic clocks 
based on Ramsey interferometry in particular. The usefulness of cold atoms is discussed, as well as their 
limits if Bose-Einstein condensates are used.  
We study as an alternative a different cold-atom regime: the Tonks-Girardeau (TG)
gas of tightly confined and strongly interacting bosons.  
The TG gas is reviewed and then generalized for two-level atoms.
Finally, we explore the combination of Ramsey interferometry and 
TG gases.  
\end{abstract}
\maketitle

\section{Introduction}
It is hardly necessary to insist on the practical importance of time
since we all live attached to a time machine, the wrist watch, which 
organizes our daily routine.    
For running the Blaubeuren Conference, a precision of one
minute is enough, but for plenty of 
activities fundamental to modern society a much more precise time keeping is necessary: banks, electric power companies, telecommunications, 
or the GPS use atomic clocks.  


A clock is a device that, in a sense, ``produces'' time by counting
stable oscillations, for example of a pendulum.
Clocks of many different types have been used along
history by different civilizations but the Earth rotation 
has been for centuries the master clock. Nevertheless, we know that the rotation speed may be affected by a series of physical phenomena, such as tidal friction, which slows down the Earth at a non-negligible pace in the geological
scale (a poor dinosaur had to rush for a sandwich in stressful days of 23 hours);
even the 
accumulation of snow on the top of the mountains in the winters of the Northern hemisphere
may affect the rotation speed -by conservation of angular momentum- making winter days indeed longer than summer days.
If we kept the definition of the second attached to the actual Earth's
rotation, we should also change the laws of physics from summer to winter, 
and add proper 
correction terms to account for tidal friction. 
This sounds very unreasonable. Instead, we could also try to correct 
for these perturbing effects ``by hand'', but unfortunately not all the perturbations are easy to predict or understand (e.g., those due to magma
motion), so corrections using other astronomical periods were explored, but they 
were very cumbersome to handle, and not stable enough anyway.  
Fortunately nature has a lot of stability to offer in the opposite direction, 
in the world of the small.     

Atomic clocks, in particular, count the oscillations of the field resonant with an atomic transition, most frequently a hyperfine transition of caesium-133
(which defines officially the second as the time required for 9192631770
periods of the resonant microwave field).   
An external quartz oscillator is locked by a servo loop to a
resonance excitation curve between two hyperfine states $|g\ra$ and $|e\ra$
so that its frequency is always adjusted to the maximum 
of the curve, and thus to the natural frequency of the selected transition.    

\begin{figure}[h!]
\begin{center}
\includegraphics[width=11cm]{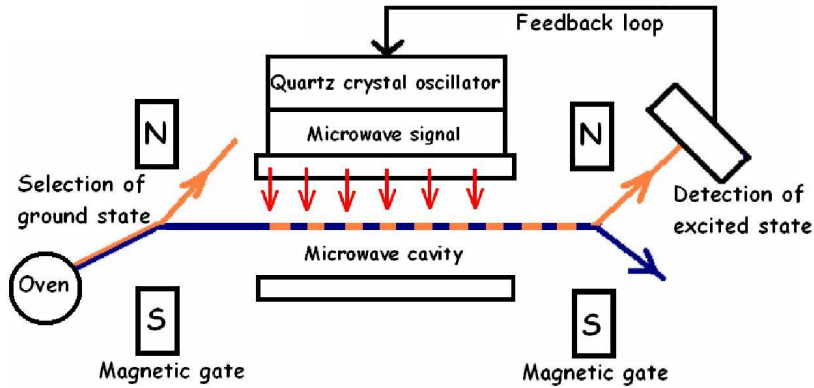}
\end{center}
\caption{\label{acl}Principle of an atomic clock based on a ``Rabi'', or one-cavity scheme. Deflection angles are greatly exaggerated. 
Actual clocks operate usually with two separated cavities (``Ramsey scheme'').}
\end{figure}

\begin{figure}[h!]
\begin{center}
\includegraphics[width=10cm]{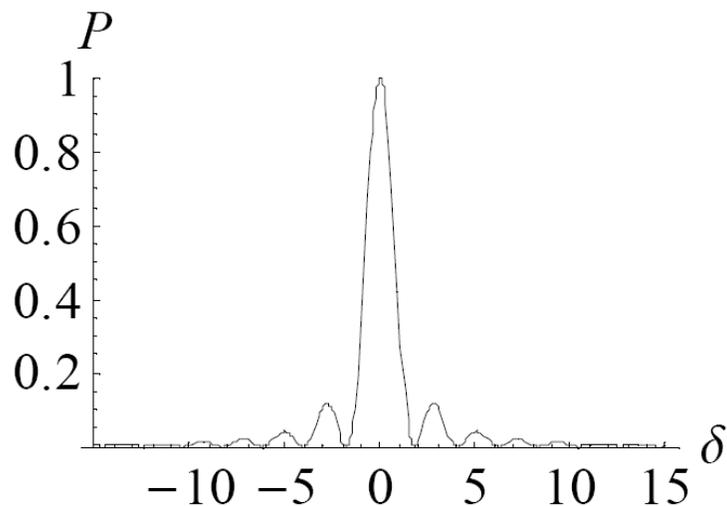}
\end{center}
\caption{\label{ep}An excitation probability curve versus detuning (arbitrary units).}
\end{figure}

In more detail, see Fig. \ref{acl}, caesium atoms are heated in an oven 
to produce a beam. 
A first magnet filters out the excited atoms so that only ground state atoms 
enter into the microwave cavity in which the field wavevector is perpendicular
to the atomic motion, and the frequency, locked to an external quartz oscillator, 
is very close to the transition frequency. This excites some atoms which are selected 
with a second magnet and later detected.
The excitation probability compared with previous 
runs at slightly different frequencies tells us how far or close is the field frequency to the transition frequency, and this information is used to modify the excitation frequency 
so that it stays as close as possible to the maximum of the excitation curve, 
see Fig. \ref{ep}.  
The clock includes the appropriate counting electronics which we shall not discuss
here.    
A very sharp peak is clearly desirable to minimize the fluctuation of the  
external oscillation frequency around the natural one. This contributes to the stability of the clock
and explains why a beam configuration is chosen: a perpendicular excitation 
with respect to the atomic motion avoids the Doppler effect and its 
associated line broadening.   
  
In actual clocks things are slightly more complicated because instead of a single cavity (Rabi scheme) there are actually two (Ramsey scheme). 
The reason for the two cavities is that they produce a quantum interference 
between two possible paths corresponding to excitation in the first 
cavity or excitation in the second, and this interference may be used to 
sharpen the resonance and to make it less dependent of  
inhomogeneities of the fields.   
The Hamiltonian describing the process  is 
\begin{equation}
H = \frac{\widehat{p}^{\,2}}{2m} - 
\hbar\delta \ket{e}\bra{e} + \frac{\hbar}{2}\Omega(\widehat{x}) (\ket{g}\bra{e} + \ket{e}\bra{g}),
\label{hami}
\end{equation}
\begin{figure}[h]
\begin{center}
\includegraphics[width=15cm]{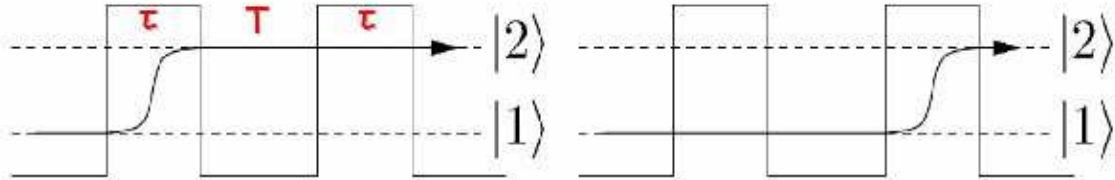}
\end{center}
\caption{\label{ram} The two main interfering paths in Ramsey's method.  
Here $|1\ra=|g\ra$, $|2\ra=|e\ra$. 
}
\end{figure}
where $\Omega$ is the Rabi frequency and $\delta$ the detuning (laser frequency minus 
transition frequency). In the semiclassical approximation the center of mass motion
is classical so that $\widehat{p}$ and $\widehat{x}$ become numbers: $p$ is  
the momentum and $x$ a linear function of time; $T$ is the flight time between the fields and $\tau$ the time at each field region,     
see Fig. \ref{ram}. Note that the second path (excitation in the second field) does not pick up any extra phase
during free flight between the fields but the first one does ($e^{i\delta T}$). 
The consequence is a final excitation probability proportional to 
$\cos^2(\delta T/2)$, 
which makes clear that longer free-flight times are desirable in order to 
achieve a narrower central fringe. 

This motivates the use of cold atoms in time-frequency metrology. 
They are slow and big $T$'s may be produced, but there are  
other advantages: first they may reduce dramatically 
velocity broadening (the averaging of the fringes
due to different velocities and times), and also 
they may lead to fundamentally new effects with coherent N-body states (e.g. entanglement has been proposed to beat quantum projection noise limit).        
 
So why do not we try to use Bose-Einstein condensates? 
Before we answer this question, let us tell a little story to show that  
``the colder the better'' is not an infallible motto.    
The Chinese made very complex water clocks centuries ago in which     
the falling water
moved a wheel with small cups. Some versions of these clocks 
were also built in Europe, but they obviously had serious problems 
in cold winters: the water froze and the clock did not work. 

As in our water clock example, the improvements associated with low velocities and narrow velocity distribution in a Bose-Einstein condensate may be 
compensated by negative effects, such as collisional shifts and 
instabilities leading to the separation of the gas cloud \cite{Cornell02,Band06}.
So BECs do not seem to make good clocks, but there are other cold-atom coherent 
regimes. In particular, the 
Tonks-Girardeau (TG) regime
of impenetrable, tightly confined Bosons subjected to ``contact'' 
interactions  \cite{Gir60,Gir65} is opposite to the BECs in some respects
and offers a priori interesting properties for metrology:    
the TG requirement of strong contact interactions, implies similarities 
between the Bosonic system and a ``dual'' system of freely moving Fermions, 
since the particle densities 
of the TG Bosons and the dual system of Fermions are actually equal, 
so we may expect that the strong interaction regime is actually an asset 
for a good clock.  
Other important feature of the TG gas is its one dimensional (1D) character.   
Olshanii showed \cite{Ols98,BerMooOls03,PetShlWal00} that when 
a Bosonic vapor
is confined in a wave guide with tight transverse trapping and 
temperature so low that the transverse vibrational excitation quantum 
$\hbar\omega_{\perp}$ is larger than available longitudinal zero point and 
thermal energies, the effective dynamics becomes one dimensional, 
and accurately described by a 1D Hamiltonian with delta-function interactions
$g_{1D}\delta(x_j-x_{\ell})$, where $x_j$ and $x_{\ell}$ are 1D longitudinal
position variables. This is the Lieb-Liniger (LL) 
model \cite{LieLin63}.
The coupling constant $g_{1D}$ can be tuned  
by varying the magnetic field (and thus the three dimensional s-wave scattering length) 
or the confinement (confinement induced resonances) \cite{Ols98,BerMooOls03}) near a Feschbach resonance; 
it is thus possible to reach the Tonks-Girardeau regime of impenetrable Bosons,
which corresponds to the $g_{1D}\to\infty$ limit of the LL model.
The first experiments were carried out in 2004 \cite{Par04,Kin04}, whereas  
the TG gas model was proposed and solved exactly in 1960 by M. Girardeau \cite{Gir60,Gir65} exploiting the similarities between the Bose gas and its dual 
ideal Fermi system mentioned above with the so called Fermi-Bose mapping.   
In actual confined gases the TG regime requires a large ratio between the chemical potential and the kinetic energy,
$\gamma=mg_{1D}/\hbar^2 n$ \cite{RT04}, $n$ being the linear density.
Note that a smaller density is favorable (to make this more intuitive, think of a rescaling of the lenghts in which the interatomic interaction is effectively closer to a delta function as the gas becomes more dilute). On the other hand we do not 
want $n$ to be too small for a clock since the total number of particles 
(and thus the signal) would also decrease.      

The recipe to construct an $N$-boson TG-wavefunction by Fermi-Bose mapping 
is as follows: 
\begin{itemize} 
\item Build a Slater determinant for ideal (free) spinless Fermions using one-particle �orbitals�
\beq
\psi_{F}(x_{1},\dots,x_{N})=\frac{1}{\sqrt{N!}}det_{n,k=1}^{N}\phi_{n}(x_{k}).
\eeq
\item  Apply the ``antisymmetric unit function" 
$\mathcal{A}=\prod_{1\leq j<k\leq N}sgn(x_{k}-x_{j})$, as 
\beq
\psi_{B}(x_{1},\dots,x_{N})=
\mathcal{A}(x_{1},\dots,x_{N})\psi_{F}(x_{1},\dots,x_{N}).
\eeq
\end{itemize}
In metrology and atomic interferometry, the tight 1D confinement
along a waveguide
has pros and cons: the absence of transversal excitation and motional branches
may lead to an increased signal, 
but the confinement is by itself problematic for frequency standard 
applications, since  
the necessary magnetic or optical interactions will in principle perturb the internal
state levels of the atom. Several schemes have been proposed
to mitigate this problem and compensate or avoid the shifts   
\cite{Cornell02,ss,Ha}, and    
we shall here assume that such a compensation is implemented.

  
To investigate the implications in Ramsey interferometry of a strongly interacting  
1D gas, it is first necessary to generalize the TG gas model 
by adding internal structure. The resulting 
idealized model remains exactly solvable 
if the collisions produce internal state and momentum exchange.  
We shall also briefly discuss the Ramsey scheme in the time domain, 
which turns out to give very similar results  
for reasonable parameters. 
\section{Two-level Tonks-Girardeau gas with exchange, contact interactions}
%
%
%
%
%
%
%
%
%
%
The first step to generalize the TG gas to one with internal states 
is to use instead of one-particle 
orbitals, one-particle ``spinors''   
\beqa
\Phi_n(x_1)=
\sum_{b=g,e}\phi^{(b)}_n(x_1)\vert b\ra, 
\eeqa
%
where $n=1,2,3...$ is a label to distinguish different spinors (it will later on 
correspond to states prepared as harmonic oscillator eigenstates of a longitudinal trap),
and $b$ may be $g$ (ground), 
or $e$ (excited) ($g$ and $e$ do not necessarily
correspond to states with definite values of the component of the electronic spin in one direction.)     
Two-particle states may also be formed as  
\beq
\Phi_{nn'}(x_1,x_2)=
\sum_{b,b'}\phi_n^{(b)}(x_1)\phi_{n'}^{(b')}(x_2)|bb'\ra, 
\eeq
and similarly for more particles. 
In $|bb'\ra$, $b$ is for particle 1 and 
$b'$ for particle 2. This will in some equations be indicated even more explicitly 
as $b_1$, $b_2$, etc.

Analogously to the steps to construct the ordinary TG gas wavefunction, 
let us first define a Fermionic state for two noninteracting  
particles with internal structure  
as a generalized Slater determinant, 
\beqa
\Psi_F(x_1,x_2)
&=&
\frac{1}{\sqrt{2}}{\rm det}_{n,j=1}^{2}\Phi_{n}(x_{j})
\nonumber\\
&=&\frac{1}{\sqrt{2}}
\left\vert\begin{array}{cc}
\Phi_1(x_1) & \Phi_1(x_2)
\\ 
\Phi_2(x_1) & \Phi_2(x_2)
\end{array}\right\vert
\nonumber
\\
&=&\frac{1}{\sqrt{2}}\sum_{b_1,b_2=g,e}
\left\vert\begin{array}{cc}
\phi_1^{(b_1)}(x_1) & \phi_1^{(b_2)}(x_2) \\
\phi_2^{(b_1)}(x_1) & \phi_2^{(b_2)}(x_2)
\end{array}\right\vert
\vert b_1 b_2\ra.
\nonumber
\eeqa
More explicitly, 
\beqa
&&2^{1/2}\Psi_F(x_{1},x_{2})=
\nonumber\\
&+&[\phi_1^{(g)}(x_1)\phi_2^{(g)}(x_2)-\phi_2^{(g)}(x_1)\phi_1^{(g)}(x_2)]|gg\ra
\nonumber\\
&+&[\phi_1^{(e)}(x_1)\phi_2^{(e)}(x_2)-\phi_2^{(e)}(x_1)\phi_1^{(e)}(x_2)]|ee\ra
\nonumber\\
&+&[\phi_1^{(g)}(x_1)\phi_2^{(e)}(x_2) -\phi_2^{(g)}(x_1)\phi_1^{(e)}(x_2)]|ge\ra
\nonumber\\
&+&[\phi_1^{(e)}(x_1)\phi_2^{(g)}(x_2) -\phi_2^{(e)}(x_1)\phi_1^{(g)}(x_2)]|eg\ra.
\label{ferm}
\eeqa
These Fermions do not interact among themselves, but they could interact with 
an external potential.
 
A Bosonic wavefunction of interacting atoms,
symmetric under $(x_i,b_i)\leftrightarrow(x_j,b_j)$ permutations 
may be now obtained 
by means of the Bose-Fermi mapping, 
$\Psi_B(x_1,x_2)=\mathcal{A}\Psi_F(x_1,x_2)$, where the 
antisymmetric unit function is 
$\mathcal{A}={\rm sgn}(x_{2}-x_{1})$.

For the sector $x_1<x_2$,    
\beqa
&&2^{1/2}\Psi_B(x_{1},x_{2})=
\nonumber\\
&+&[\phi_1^{(g)}(x_1)\phi_2^{(g)}(x_2)-\phi_2^{(g)}(x_1)\phi_1^{(g)}(x_1)]|gg\ra
\nonumber\\
&+&[\phi_1^{(e)}(x_1)\phi_2^{(e)}(x_2)-\phi_2^{(e)}(x_1)\phi_1^{(e)}(x_2)]|ee\ra
\nonumber\\
&+&[\phi_1^{(g)}(x_1)\phi_2^{(e)}(x_2)]|ge\ra -[\phi_2^{(e)}(x_1)\phi_1^{(g)}(x_2)]|eg\ra
\nonumber\\
&+&[\phi_1^{(e)}(x_1)\phi_2^{(g)}(x_2)]|eg\ra
-[\phi_2^{(g)}(x_1)\phi_1^{(e)}(x_2)]|ge\ra.  
\label{bosons}
\eeqa
In the complementary sector $x_1>x_2$ we would get the same form except for a global 
minus sign. 
The resulting Bosonic state is discontinuous at contact so it cannot represent noninteracting Bosons. 
The generalization to $N$-particles is straightforward \cite{pra}.   
 
To understand the physical meaning of the contact interactions 
implicit in Eq. (\ref{bosons}),  
let us now introduce Pauli operators,
\beqa
\sigma_{X}&=&|g\ra\la e|+|e\ra\la g|,
\nonumber\\
\sigma_Y&=&i(|g\ra\la e|-|e\ra\la g|),
\nonumber\\
\sigma_Z&=&|e\ra\la e| -|g\ra\la g|,
\eeqa
and the corresponding 3-component operator 
$\hat{\mathbf{S}}_j={\sigma}_j/2$ for particle $j$
analogous to the spin-$1/2$ angular momentum operator.  
For two particles $\hat{\mathbf{S}}=\hat{\mathbf{S}}_1+\hat{\mathbf{S}}_2$,
and $\hat{\mathbf{S}}^2$ has eigenvalues $S(S+1)$ with $S=0$ and $S=1$
corresponding to singlet and triplet subspaces. These subspaces are spanned by 
$|-\ra\equiv(|eg\ra-|ge\ra)/\sqrt{2}$
and $\{|gg\ra,|ee\ra,|+\ra\equiv(|eg\ra+|ge\ra)/\sqrt{2}\}$
respectively.  

Assume now the Hamiltonian
\begin{equation}\label{Fermi Hamiltonian_2}
\hat{H}_{\rm{coll}}=-\frac{\hbar^2}{2m}\sum_{j=1}^{2}\partial_{x_j}^{2}
+v_s(x_{12})\hat{P}_{12}^s
+v_{t}(x_{12})\hat{P}_{12}^{t}.
\end{equation}
Here $x_{12}=x_1-x_{2}$, 
$\hat{P}_{12}^s=|-\ra\la -|=\frac{1}{4}-\hat{\mathbf{S}}_1\cdot\hat{\mathbf{S}}_{2}$ 
is the projector onto the  subspaces of singlet functions, and 
$\hat{P}_{12}^{t}=|gg\ra\la gg|+|ee\ra \la ee|+|+\ra\la +|$
%
is a  projector onto the triplet subspace.   
\begin{figure}
\begin{center}
\includegraphics[width=11cm]{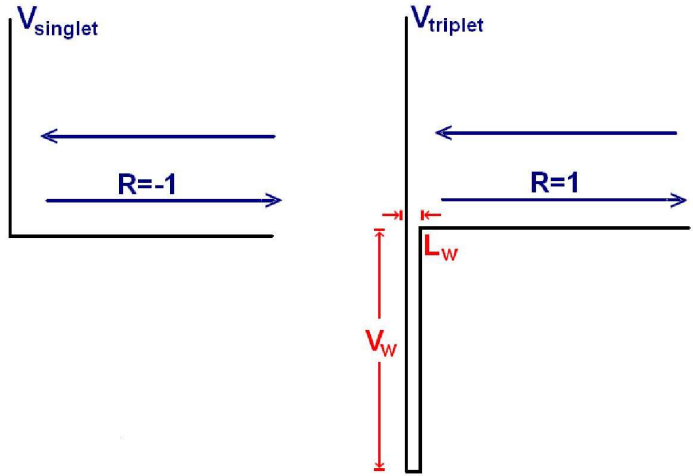}
\end{center}
\vspace*{-1.5cm}
\caption{\label{label}Collisions for triplet and singlet potentials.}
\end{figure}
      
The internal Hilbert space can be written as the sum of singlet and triplet 
subspaces as $\mathcal{H}_s\oplus\mathcal{H}_t$.  
Suppose that the reflection amplitude for relative motion 
in such representation takes the values $+1,-1$ 
in singlet and triplet subspaces respectively. 
The particles are impenetrable 
and these values correspond to a hard wall potential $v_t$ at $x_{12}=0$
for all collisions in triplet channels $|gg\ra\to|gg\ra$, $|ee\ra\to|ee\ra$ or $|+\ra\to |+\ra$,    
whereas $v_s$ has, in addition to the hard wall at $x_{12}=0$,  
a well of width $l$ and 
depth $V$,  
so that the reflection amplitude becomes $R=+1$ in the limit in which the well is made infinitely narrow and the well 
infinitely deep, keeping $(2mV_w/\hbar^2)^{1/2} L_w=\pi/2$ \cite{GirOls03,GirOls04,GirNguOls04,CheShi98}.
(In the Fermionic Tonks-Girardeau gas the well applies to the
triplet subspace and not to the singlet subspace as here.)  

\begin{figure}
\includegraphics[width=7cm,angle=0]{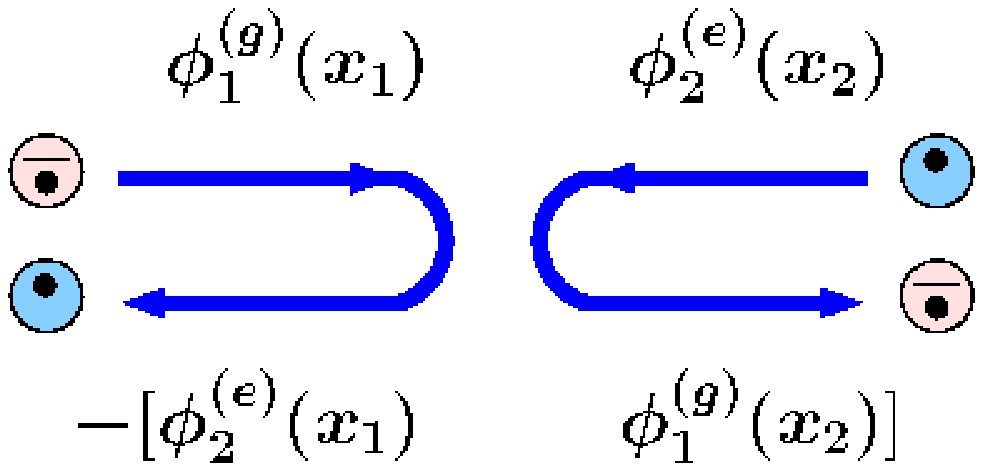} 
\hspace*{1.8cm}
\includegraphics[width=7cm,angle=0]{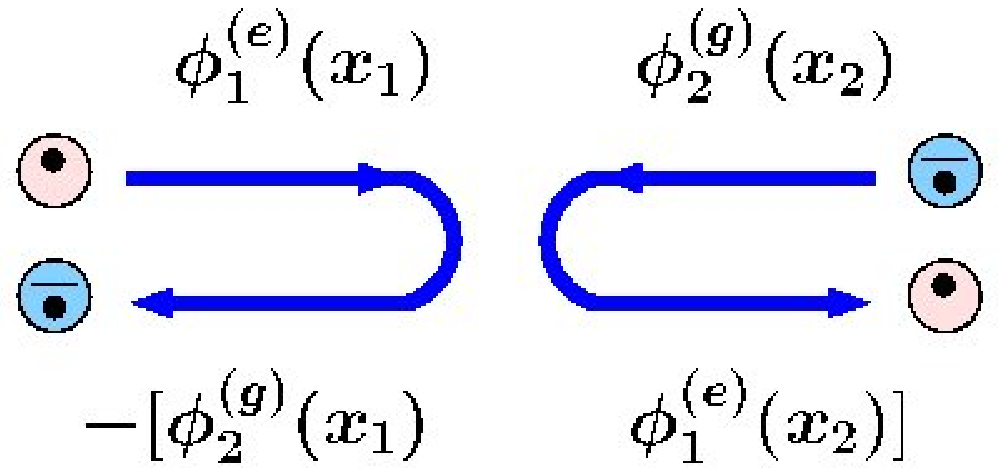} 
  \caption{Diagrammatic representation of the collisions for particles with   
different internal states in the sector $x_1<x_2$ (the last two terms in Eq. (\ref{bosons})):
the particles interchange their 
momentum and internal state picking up an additional phase (minus sign).
} \label{coll}
\end{figure}

If we translate the above into the $g,e$-basis and for the sector $x_1<x_2$ this implies that in all collisions between atoms with well defined incident momenta, 
they interchange their momenta 
(the relative momentum changes sign), as well as their internal state,
with the outgoing wave
function picking up a minus sign because of the hard-core reflection, as shown in Fig. \ref{coll}. 
For $x_1<x_2$ and equal internal states such collision is represented by
\beq
e^{ikx_1}e^{ik'x_2}|bb\ra-[e^{ik'x_1}e^{ikx_2}]|bb\ra,
\label{diag}
\eeq
whereas for $b\ne b'$,
\beq
e^{ikx_1}e^{ik'x_2}|bb'\ra-[e^{ik'x_1}e^{ikx_2}]|b'b\ra.
\label{ndiag}
\eeq
For equal internal states the spatial part vanishes at contact, $x_1=x_2$,
whereas in the non-diagonal case it does not, but in Eq. (\ref{ndiag}) only the 
``external region'' is considered, 
disregarding the infinitely narrow well region.    
%
%

For an implementation 
of these contact interactions,  
undesired inelastic collisions should be suppressed 
by confinement at low collision energies  
\cite{YB07}, which will also produce 
effective triplet reflection coefficients close to $-1$, independently
of the internal state. 


For two Bosons 
in the singlet subspace, 
the space wavefunction is antisymmetric, so that s-wave scattering is forbidden;  
therefore the interactions are governed to leading order by a 3D p-wave scattering amplitude 
and can be enhanced by a 1D odd-wave confinement-induced  
Feschbach resonance (CIR), which allows in principle to engineer $v_s$
and achieve a strong attraction as required above.   

\section{The Ramsey interferometer with a TG gas}
\begin{figure}
\begin{center} 
\includegraphics[angle=0,width=12 cm]{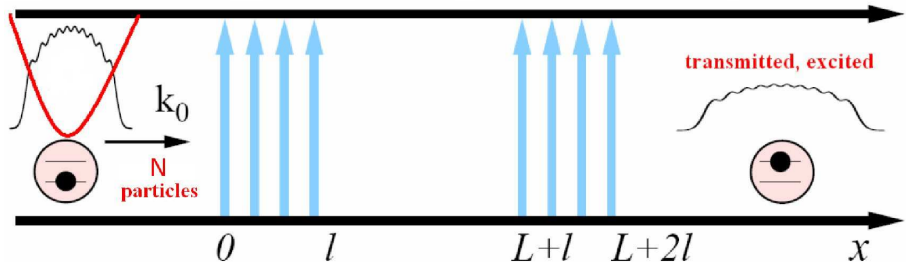}
\vspace*{-2cm}
\caption{\label{setup} Schematic setup for Ramsey interferometry 
of guided atoms in the spatial domain.
The atoms are prepared in the ground state and the probability of excitation 
is measured after passing the two fields.}
\end{center}
\end{figure}

%
%
%
%
%
%
%
We consider as the initial state for the Ramsey 
experiment $N$ two-level atoms in the Tonks-Girardeau regime,
confined in their ground internal states in a harmonic trap of frequency $\om$.
The cloud is released by switching off the trap along the x-axis at time $t=0$ ($\om=0$ for $t>0$);  
a momentum kick $\hbar k_0$ is also applied, so that it moves  
towards the two oscillating fields localized between $0$ and $l$ and between $l+L$ and $2l+L$ (Fig.~\ref{setup}).
The central initial position of the harmonic trap $x_0<0$ is such that  $|x_0|>>\delta_N=[(N+1/2)\hbar/(m\omega)]^{1/2}$, to avoid the overlap 
with the spatial width (root of the variance) of the highest occupied state.       
   
In an oscillating-field-adapted interaction picture 
(which does not affect the collisional Hamiltonian)
and using the Lamb-Dicke, dipole, and rotating-wave approximations 
the Hamiltonian is given, for each of the particles,
by Eq. (\ref{hami}),    
%
%
For the explicit $x$ dependence of $\Omega(x)$ we assume $\Omega(x) = \Omega$ for $x\in [0,l]$ and $x \in [l+L,2l+L]$ and zero elsewhere. We include the interparticle interactions implicitly by means of the TG wave function 
and its boundary conditions at contact.    
%
%

%

%
The Ramsey pattern is defined by the dependence on the detuning 
of the probability of excited atoms after the interaction with the two field regions. 
For the TG gas it follows that the total excitation probability is obtained as the 
average of the excitation probabilities for each spinor,  $P_{e}=\frac{1}{N}\sum_{n=1}^{N}p_{n}^{(e)}$.   
Once a particle incident from the left and prepared in the state 
$e^{ik_0(x-x_0)}\phi_{n}(x-x_0)|g\ra$ at $t=0$ 
has passed completely through both fields, the probability amplitude for it to be in the 
excited state is 
\beqa
\phi_n^{(e)}(x,t)=\frac{1}{\sqrt{2\pi}}\int\d k e^{iqx-ik^2\hbar t/(2m)}T_{ge}(k)\tilde{\phi}_n(k),
\eeqa
where $\tilde{\phi}_n(k)$ is the wavenumber representation of the kicked $n$-th harmonic 
eigenstate, 
\beqa
\tilde{\phi}_n(k)&=&\frac{(-i)^n}{\sqrt{2^n n!}}\left(\frac{2\delta_0^2}{\pi}\right)^{1/4}
e^{-\delta_0^2(k-k_0)^2}e^{-ik x_0}H_n[\sqrt{2}\delta_0(k-k_0)],
\eeqa
the momentum in the excited state is $q = \sqrt{k^2+2m\Delta/\hbar}$, the spatial width of the $n=0$ state is $\delta_0=[\hbar/(2 m\om)]^{1/2}$, $H_n$ the Hermite polynomials, and $T_{ge}$ is the
``double-barrier'' transmission amplitude for the 
excited state corresponding to atoms incident in the ground state (the excited state probability 
for monochromatic incidence in the ground state is 
$\frac{q}{k}|T_{ge}|^2$). 
A fully quantum treatment of $T_{ge}$ can be found in \cite{SeiMu-EPJD}.
We have done numerical simulations for $l=1$ cm, $L=10$ cm, 
$N=10$, and $v_0=1$ cm/s. 
The variation of the excitation probability for
different harmonic eigenstates is negligible, and  
the curves for the central fringe are indistinguishable from the semiclassical 
result of Ramsey (classical motion for the center of mass,
uncoupled from the internal levels), 
\begin{eqnarray}
\label{eq:Ramsey_limit}
P_{e,scl}(\Delta) &=& \frac{4\Omega^2}{\Omega'^2} \sin^2\left(\frac{\Omega' \tau}{2} \right) 
\bigg[\cos\left(\frac{\Omega' \tau}{2} \right) \cos\left(\frac{\Delta T}{2}  \right) 
-\frac{\Delta}{\Omega'} \sin\left(\frac{\Omega' \tau}{2} \right) \sin\left(\frac{\Delta T }{2}  \right) 
\bigg]^2,
\end{eqnarray}
where $\tau=l/v_0$, $T=L/v_0$, and $\Omega'=(\Omega^2+\Delta^2)^{1/2}$. 

The broadening of the central fringe by increasing $n$ due to the
momentum broadening of vibrationally excited states  
is quite small for the few-body states
of our calculations, $N=10$, which is in fact of the order of
current experiments with TG gases ($N\approx 15, 50$ in \cite{Par04,Kin04}), 
see \cite{pra} for details.  

We should also examine the quantum projection noise due to the fact that
only a finite number of measurements are made to determine
the excitation probability. 
The error to estimate the atomic frequency from the Ramsey pattern depends on the ratio 
\cite{Itano93}
\beq
r=\frac{\Delta S_Z}{|\partial \la S_Z\ra/\partial\delta|},
\label{r}
\eeq
which we have calculated at half height of the central interference peak. 
Even though $r$ is  theoretically smaller for the TG gas than for independent
atoms, 
for reasonable parameters the ratio $r$ essentially 
coincides with that for freely moving, uncorrelated particles \cite{Itano93} and, for $L>>l$  it gives $1/(T\sqrt{N})$ for all $\delta$ \cite{pra}.

\section{Ramsey interferometry in the time domain for guided atoms}
\begin{figure}[t]
\begin{center}
\includegraphics[height=3cm]{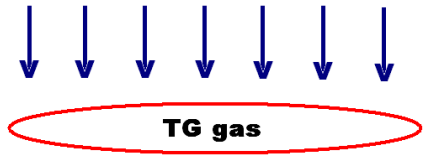}
\caption[]{Schematic setup for Ramsey interferometry in time domain. 
The TG gas is confined in a cigar-shaped trap and illuminated by a laser in $y$-direction.}
\label{cigar_trap}
\end{center}
\end{figure}

An alternative to the previous set-up is the separation of the fields 
in time rather than space but, at variance with the usual procedure, 
keeping the gas confined transversally at all times as required for the $1D$ regime of the 
TG gas, Fig. \ref{cigar_trap}. 
This configuration is thus similar to the one used in ion
trap clocks. 
Because of the tight confinement the transverse vibrational excitation is 
negligible so that the Ramsey pattern is given by the standard semiclassical 
expression irrespective of the value of $n$. The whole TG gas therefore
produces the usual Ramsey pattern (\ref{eq:Ramsey_limit}).
The advantage with respect to trapped ions is the augmented signal, and 
a disadvantage would be the need to take care of, and possibly correct for, 
perturbing effect of the confinement on the levels of the neutral atoms.  
         
%
%

%

%
%
%
%
%
%

\section{Summary and discussion}

Our model of a generalized Tonks-Girardeau regime suggests 
that very cold atoms may provide good atomic clocks if strongly confined. 
The advantages are:  
a very large time between field interactions,  
no collisional shifts,   
no velocity broadening, and  
no instabilities as in BECs. 
More work is needed to ascertain the role of strongly interacting gases 
in interferometry and time-frequency metrology:  
to go from models to actual atoms,
and to compensate or eliminate shifts due to confinement forces.

\begin{acknowledgments}
We acknowledge discussions with M. Girardeau.
This work has been supported by Ministerio de Edu\-ca\-ci\'on y Ciencia (FIS2006-10268-C03-01; FIS2005-01369) and UPV-EHU (00039.310-15968/2004).
S. V. M. acknowledges a research visitor Ph. D. student fellowship  
by the Ministry of Science, Research and Technology of Iran.
A. C. acknowledges financial support by the Basque Government (BFI04.479).
\end{acknowledgments}

\end{document}